\documentclass[arxiv]{melba}

\usepackage{mwe} 

\usepackage{amsmath,amsfonts}

\usepackage{subcaption}
\usepackage{float}

\melbaid{}  
\doi{10.59275/j.melba.2026-}
\melbaauthors{Nohel and Valek et al.}  
\email{Dostal.Marek@fnbrno.cz, xnohel04@vutbr.cz}
\volume{2026}
\firstpageno{1}  
\melbayear{2026}  
\datesubmitted{yyyy-m1-d1}  
\datepublished{yyyy-m2-d2}  

\melbaspecialissue{MICCAI Open Data 2026 x MELBA}
\melbaspecialissueeditors{AT, MS et al.}

\ShortHeadings{Spinal-Multiple-Myeloma-SEG: A Dual-Energy CT Dataset Extended with Trabecular Bone Segmentation of Lumbar Vertebrae}{Nohel and Valek et al.}

\title{Spinal-Multiple-Myeloma-SEG: A Dual-Energy CT Dataset Extended with Trabecular Bone Segmentation of Lumbar Vertebrae}

\author{
    \firstname Michal \surname Nohel\aff{1,2}\orcid{0000-0002-2679-2160},
    \firstname Vlastimil \surname Valek\aff{3,4}\orcid{0000-0001-6851-9595},
    \firstname Katerina \surname Krejci\aff{1}\orcid{0009-0009-5817-4840},
    \firstname Roman \surname Jakubicek\aff{1}\orcid{0000-0003-4293-260X},
    \firstname Marek\surname Dostal\aff{3,5}\orcid{0000-0003-1740-9227},
    \firstname Jiri \surname Chmelik\aff{1}\orcid{0000-0001-9950-6279}
}
\affiliations{
    \num 1 \addr Department of Biomedical Engineering, Faculty of Electrical Engineering and Communication, Brno University of Technology, Brno, Czech Republic \\
    \num 2 \addr Department of Deputy Director for Science and Research, University Hospital \& Faculty of Medicine, Ostrava, Czech Republic \\ 
	\num 3 \addr Department of Radiology and Nuclear Medicine, University Hospital Brno, Brno, Czech Republic \\
	\num 4 \addr Department of Radiology and Nuclear Medicine, Faculty of Medicine, Masaryk University, Brno, Czech Republic \\
    \num 5 \addr Department of Biophysics, Masaryk University, Brno, Czech Republic
}

\abstract{
We present an extension of the publicly available \textit{Spinal-Multiple-Myeloma-SEG} dataset, a dual-energy CT imaging resource for multiple myeloma research. The purpose of this dataset is to enable voxel-wise analysis of vertebral bone microstructure by adding expert-validated segmentation of the trabecular compartment of lumbar vertebrae.
The dataset consists of 72 dual-energy CT examinations from 67 adult patients (mean age 66 years, range 48--85; 36\% female), acquired retrospectively using a dual-layer dual-energy CT system. It includes conventional CT, virtual monoenergetic images, and calcium-suppressed reconstructions, along with structured clinical metadata. The data are provided in DICOM format, while segmentation masks are available in both NIfTI and DICOM-SEG formats.
The primary intended applications include trabecular bone segmentation, quantitative bone mineral density-related analysis, and development of deep learning models for vertebral anatomy and disease-affected bone structures in multiple myeloma. The dataset supports both segmentation and multimodal learning tasks in pathological and non-pathological bone.
Initial trabecular segmentation masks were generated using a pretrained nnU-Net model and subsequently refined through manual expert correction and radiological quality control, ensuring anatomical consistency.
The original dataset is publicly available via TCIA (\href{https://doi.org/10.7937/k4qv-hh78}{https://doi.org/10.7937/k4qv-hh78}), while the trabecular segmentation extension (Version 2) is released through Zenodo (\href{https://doi.org/10.5281/zenodo.21628232}{https://doi.org/10.5281/zenodo.21628232}) under the CC BY 4.0 license. The Zenodo release provides immediate public access to the segmentation masks and will be additionally incorporated into the official TCIA collection after completion of the curation process.
}


\begin{document}

\twocolumn[\maketitle]

\section{Background}
    \enluminure{M}{ultiple} myeloma (MM) is a malignant plasma cell disorder characterized by clonal proliferation of plasma cells in the bone marrow, leading to abnormal production of monoclonal immunoglobulins and progressive skeletal destruction. One of the characteristic clinical manifestations of MM is the development of osteolytic bone lesions, which are associated with pain, neurological complications, and pathological fractures, particularly in the axial skeleton \citep{VandeDonk2021, Cowan202221}. Early and accurate detection of skeletal involvement is essential for disease staging, treatment planning, and monitoring of therapeutic response. These findings can also help detect the progression of precursor conditions such as monoclonal gammopathy of undetermined significance (MGUS) and smoldering multiple myeloma (SMM) to manifesting disease (MM) \citep{Zorlu202574, Rajkumar202295}.

    Computed tomography (CT) plays a central role in assessing bone disease in MM due to its high spatial resolution and sensitivity to detect cortical bone destruction. In recent years, dual-energy CT (DECT) has further improved musculoskeletal imaging by enabling material decomposition techniques, such as virtual monoenergetic imaging and calcium suppression, which enhance the visibility of the lesion and improve the differentiation between healthy and pathological bone tissue \citep{Forghani2017_part1, Forghani2017_part2, Liang2024108}. Despite these advantages, publicly available datasets focused on DECT imaging of MM with voxel-wise annotations remain extremely limited.

    Existing datasets either focus on spinal metastases from multiple primary tumors, lack lesion-level segmentation, or do not provide spectral CT information \citep{Afnouch2023, Spine-Mets-CT-SEG, CMB-MML, Edelmers2024}. This creates a gap for the development and objective evaluation of machine learning methods for the detection and segmentation of MM-specific lesions.

    Vertebral bone segmentation, including trabecular bone, has previously been investigated in healthy subjects using deep learning approaches trained on annotated CT datasets \citep{Zhang2023, Zhang2024}. However, these methods have not been validated in patients with multiple myeloma, where pathological bone remodeling and focal osteolytic lesions introduce additional challenges for automated segmentation.
  
    We have previously published a related dataset \textit{Spinal-Multiple-Myeloma-SEG (Version 1)} on The Cancer Imaging Archive (TCIA), which focuses on voxel-wise segmentation of spinal multiple myeloma lesions in dual-energy CT images \citep{Nohel2026SpinalMM}. The dataset is publicly available at \href{https://doi.org/10.7937/k4qv-hh78}{TCIA repository} and provides multiple CT reconstructions together with expert-annotated segmentation masks of vertebrae and focal myeloma lesions, allowing the development and evaluation of machine learning methods for the lesion detection and the disease characterization. However, it does not include a detailed annotation of trabecular bone structures within vertebral bodies, which is addressed in the present extension.

    In this work, we extend this dataset by introducing voxel-wise annotations of trabecular bone structures in the lumbar spine, enabling more detailed analysis of vertebral microstructure in addition to focal lesion segmentation.

\section{Summary}
\begin{figure*}[ht]
		\centering
        \begin{subfigure}[b]{0.3\textwidth}
            \centering
            \includegraphics[width=\textwidth,trim={0 225 0 225},clip]{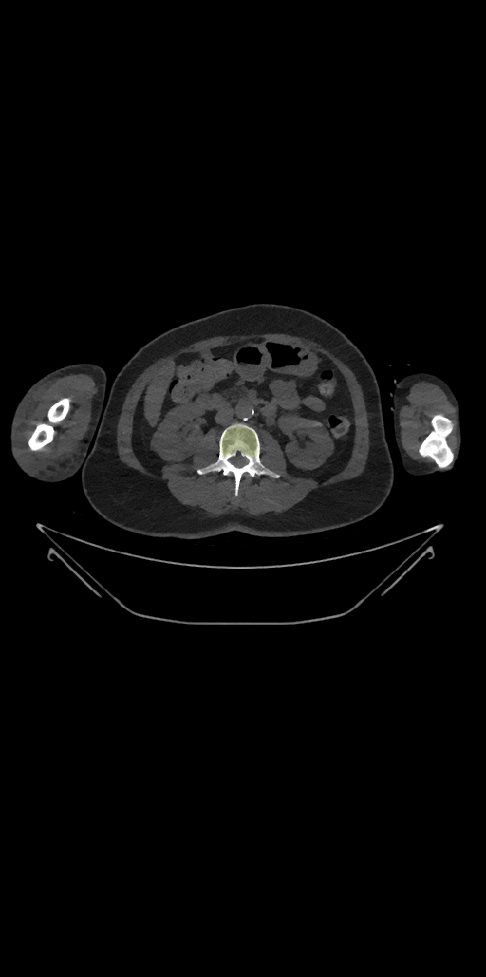}
            \caption{}
            \label{subfig:axial}
        \end{subfigure}
        \hfill
        \begin{subfigure}[b]{0.3\textwidth}
            \centering
            \includegraphics[width=\textwidth,trim={0 200 0 250},clip]{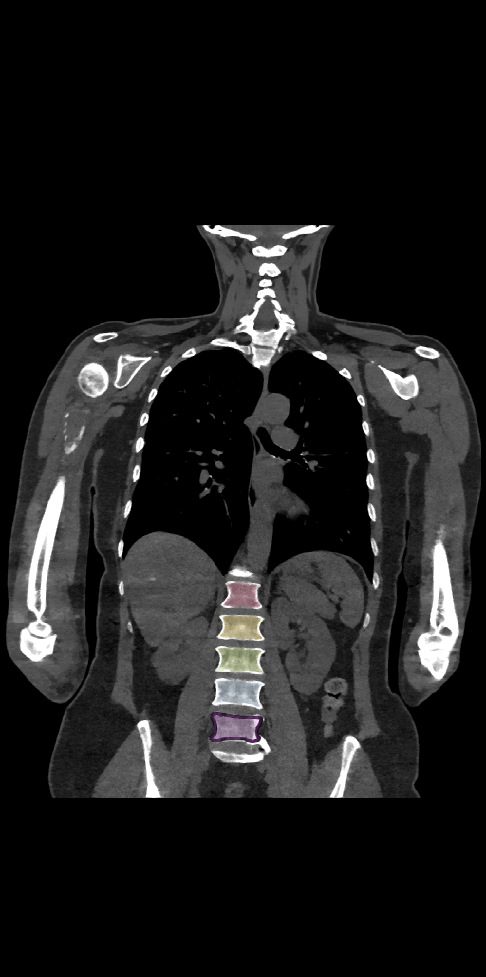}
            \caption{}
            \label{subfig:coronal}
        \end{subfigure}
        \hfill
        \begin{subfigure}[b]{0.3\textwidth}
            \centering
            \includegraphics[width=\textwidth,trim={0 200 0 250},clip]{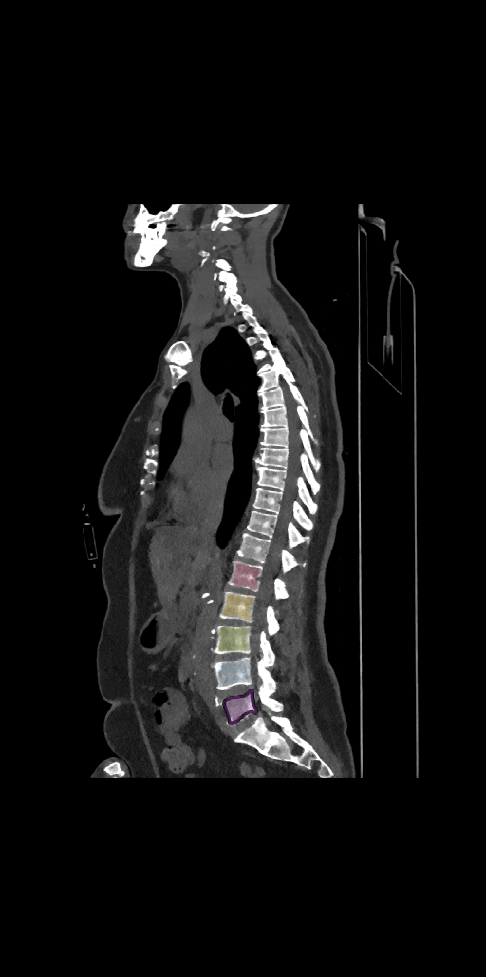}
            \caption{}
            \label{subfig:sagittal}
        \end{subfigure}
        
		\caption{Example of one patient scan. Overlay of 40 keV Virtual Monoenergetic Image (VMI) with segmentation masks of lumbar trabecular bones. Each colour represents one lumbar vertebra. Axial (\subref{subfig:axial}), coronal (\subref{subfig:coronal}), sagittal (\subref{subfig:sagittal}) slice of whole data overview.}
        
        \label{fig:data_overview}
	\end{figure*}

    The vision of this dataset is to enable robust and clinically relevant machine learning methods for quantitative analysis of skeletal involvement in MM using DECT imaging. Although existing resources primarily focus on lesion detection or vertebral segmentation, there is a lack of publicly available datasets that allow a detailed assessment of vertebral microstructure in pathological bone. The objective of this work is to provide a high-quality, expert-validated dataset that extends lesion-level annotations with voxel-wise segmentation of trabecular bone in lumbar vertebrae, thereby supporting research on bone quality assessment, disease progression analysis, and improved computer-aided diagnosis systems. In this work, we focus exclusively on voxel-wise segmentation of the trabecular compartment of the lumbar vertebral bodies, providing detailed annotations of vertebral microstructure in patients with MM. An example of a patient overview is given in Fig.~\ref{fig:data_overview}.

    The dataset is designed for use in machine learning and deep learning applications. All imaging data are provided in the DICOM format, while segmentation masks are available in both NIfTI and DICOM-SEG formats, ensuring compatibility with commonly used medical image analysis pipelines. The dataset includes multiple standardized reconstructions (conventional CT, virtual monoenergetic images, and calcium-suppressed images), enabling multimodal input strategies. In addition, structured clinical metadata are provided in tabular TSV format. The dataset is organized in a consistent hierarchical structure at the patient-level and is accompanied by a comprehensive description of the dataset and usage notes. Owing to its standardized data formats and annotation structure, the dataset is AI-ready and can be directly used for training, validation, and benchmarking of machine learning models.

    To use the dataset, users require standard medical image processing and machine learning tools capable of handling DICOM and NIfTI formats. Recommended software includes ITK-SNAP \citep{Yushkevich2006}, MITK \citep{Wolf200455}, or other appropriate tools. No proprietary software is required, and all data are distributed under a CC-BY 4.0 license, ensuring full accessibility for academic and clinical research purposes.

\section{Discussion}   
    This work represents an extension of the publicly available \textit{Spinal-Multiple-Myeloma-SEG} dataset, which is available through The Cancer Imaging Archive (TCIA) and has been previously described in \textit{Scientific Data} \citep{Nohel2026SpinalMM, Nohel_2026_SciData}. The original dataset provides voxel-wise segmentation of spinal multiple myeloma lesions in dual-energy CT images.
    
    In the present extension, we additionally provide voxel-wise annotations of the trabecular compartment of the lumbar vertebral bodies, enabling analysis of the vertebral microstructure in patients with MM.

    The presented dataset provides a unique combination of dual-energy CT imaging, expert-validated voxel-wise annotations of spinal multiple myeloma lesions, segmentation of vertebral structures, and newly introduced segmentation of the trabecular compartment of the lumbar vertebrae. The inclusion of full spine segmentation masks from the previously released TCIA dataset further enables precise anatomical localization and supports vertebra-wise and region-specific quantitative analysis.

    Compared to existing publicly available resources, the dataset includes spectral CT reconstructions (virtual monoenergetic and calcium-suppressed images), allowing advanced multimodal analysis of bone structure and pathology. The dataset is further strengthened by its relatively large cohort size for a single-center MM DECT study, consistent acquisition protocol, and availability of both imaging and structured clinical metadata.

   The initial segmentation masks were generated using a pretrained deep learning model trained on an external dataset, which was subsequently applied to the present data. The predictions were then manually inspected, corrected, and refined by a radiologist to ensure anatomical accuracy and consistency. This semi-automatic annotation workflow enabled efficient generation of high-quality, expert-validated ground truth suitable for machine learning applications.

    This dataset has several limitations. First, it originates from a single institution and a single CT scanner, which may limit generalization across different imaging systems and populations. Second, trabecular bone annotations are limited to the lumbar vertebrae (L1--L5 / L6), while other spinal regions and skeletal sites are not included.

    The dataset may be subject to selection bias, as only patients with detectable spinal lesions and available spectral CT imaging were included. This may result in an overrepresentation of more advanced disease stages. In addition, the cohort reflects a Central European population, which may limit demographic diversity. Another potential bias arises from the exclusion of cases with severe imaging artifacts.

    Users should be aware that the dataset is intended for research purposes only and not for direct clinical deployment. Models trained on this dataset should be validated on external multi-institutional cohorts before clinical translation. Particular caution should be taken when generalizing results across different CT vendors, acquisition protocols, and patient populations.

    Due to the substantially smaller volume of pathological regions compared with the trabecular bone compartment, users should carefully consider class imbalance during model training. Appropriate strategies, such as loss weighting, balanced sampling, or task-specific evaluation metrics, may be required depending on the intended application.

    Future work will focus on extending the dataset with longitudinal follow-up examinations to enable temporal analysis of disease progression and response to treatment. In addition, efforts will be directed toward expanding the dataset in a multi-center setting to improve population diversity, scanner variability, and overall robustness of AI models trained on these data.
    
    Existing datasets such as BM-Seg \citep{Afnouch2023}, Spine-Mets-CT-SEG \citep{Spine-Mets-CT-SEG}, and other spinal metastasis resources provide annotated lesion but lack dual-energy CT information and trabecular microstructural labeling. Similarly, the Cancer Moonshot Biobank Multiple Myeloma collection \citep{CMB-MML} includes multi-modal imaging data, but does not provide voxel-wise annotations for lesion or bone microstructure analysis. In contrast, the presented dataset uniquely combines DECT-based imaging, full spine segmentation, lesion segmentation, and trabecular bone segmentation within a single cohesive resource for multiple myeloma research.

\section{Resource Availability}
    \subsection{Summary statement}
    Multiple myeloma is a relatively rare hematologic malignancy with significant skeletal involvement, but publicly available imaging datasets remain limited, particularly for dual-energy CT with voxel-wise annotations. We present an extension of an existing TCIA dataset by adding expert-validated segmentation of the trabecular compartment of lumbar vertebrae. This enables a quantitative assessment of vertebral bone structure, including imaging-based biomarkers such as bone mineral density (BMD). The resource supports the development of robust machine learning models for the analysis of bone and lesions under both healthy and pathological conditions.

    \subsection{Data/Code Location}
    The original \textit{Spinal-Multiple-Myeloma-SEG} dataset (Version 1) is publicly available through The Cancer Imaging Archive (TCIA) and can be accessed via the DOI: \url{https://doi.org/10.7937/k4qv-hh78} 

    The trabecular bone segmentation extension introduced in this work (\textit{Spinal-Multiple-Myeloma-SEG}, Version 2) is publicly available through Zenodo under the DOI: \url{https://doi.org/10.5281/zenodo.21628232} \citep{nohel_2026_21628232}. The Zenodo record represents the permanent public release of the extension and provides the segmentation masks together with the accompanying dataset description files.
    
    The Version 2 extension is currently undergoing the TCIA curation process and will be additionally integrated into the official \textit{Spinal-Multiple-Myeloma-SEG} collection after completion of the TCIA review and ingestion workflow.
    
    No additional approval is required for the download. All users must comply with the TCIA data usage policy. A supporting code repository for data visualization and preprocessing is available at \href{https://github.com/MISAG-BUT/Spinal-Multiple-Myeloma-SEG}{https://github.com/MISAG-BUT/Spinal-Multiple-Myeloma-SEG}.
        
    The pretrained nnU-Net model used for the segmentation of the trabecular compartment of the lumbar vertebrae is publicly available on the Zenodo repository (\url{https://doi.org/10.5281/zenodo.11245116}) \citep{curillova_2024_11245116}. The model was trained in the LumVBCanSeg dataset \citep{Zhang2023} for cancellous bone segmentation in CT images of the lumbar spine, and its development is described in conference publication \cite{Nagyova2024}. 

    \subsection{Potential Use Cases}
    The dataset is intended for research in medical image analysis and machine learning, with a particular focus on the quantitative assessment of vertebral trabecular bone using dedicated segmentation masks. The provided trabecular compartment annotations enable BMD-related analyses and support the development of models for structural characterization of vertebral bone.

    In addition, the dataset facilitates the training of robust deep learning models for vertebra and bone segmentation in both healthy and pathological conditions, including cases affected by osteolytic lesions in multiple myeloma. The inclusion of dual-energy CT reconstructions further enables multimodal learning approaches for improved tissue characterization and disease-specific analysis.

    \subsection{Licensing}
	The dataset is released under the Creative Commons Attribution 4.0 International License (CC BY 4.0). The license allows for unrestricted use, distribution, and reproduction in any medium, provided the original authors and source are credited properly.

    \subsection{Ethical Considerations}
	The study was approved by the Institutional Review Board of the University Hospital Brno under application number NU23J-08-00027. Data were retrospectively collected from routine clinical practice. All patients provided their informed consent for the use of their imaging data for research purposes. Questions regarding data governance, ethical considerations, or responsible use, should be directed to the author (Marek Dostal, dostal.marek@fnbrno.cz).

    The dataset includes adult patients with multiple myeloma and does not contain pediatric data or other vulnerable populations. All data were pseudonymized prior to release, including the removal of direct identifiers. In addition, defacing was performed as part of the TCIA curation process to further ensure anonymization.

    The data are classified as pseudonymized medical imaging data. No identifiable personal information is included in the released dataset.

\section{Methods}
    \subsection{Data Details}
    A retrospective cohort study was conducted using DECT examinations acquired in the Department of Radiology and Nuclear Medicine, University Hospital Brno, between 2020 and 2023. The dataset includes 67 patients with a confirmed diagnosis of symptomatic multiple myeloma according to the criteria of the International Myeloma Working Group (IMWG) \citep{Hillengass2019}. In total, 72 CT studies were included, with five patients contributing follow-up examinations.

    The cohort consists of adult patients, with a mean age of 66 years (range 48–85 years) and 36\% female patients. Only patients with focal osteolytic lesions in the spine and available spectral CT data were included. The exclusion criteria comprised the absence of spectral imaging, spinal lesions, and severe imaging artifacts (e.g., metal fixation) preventing a reliable evaluation.

    A comprehensive description of the dataset, including extended clinical metadata, annotation protocols, and validation procedures, is provided in a companion manuscript accepted for publication in \textit{Scientific Data} \cite{Nohel_2026_SciData}.

    \subsection{Methods Used for the Data Creation}

    \begin{figure*}[ht]
		\centering
        \begin{subfigure}[b]{0.23\textwidth}
            \centering
            \includegraphics[width=\textwidth]{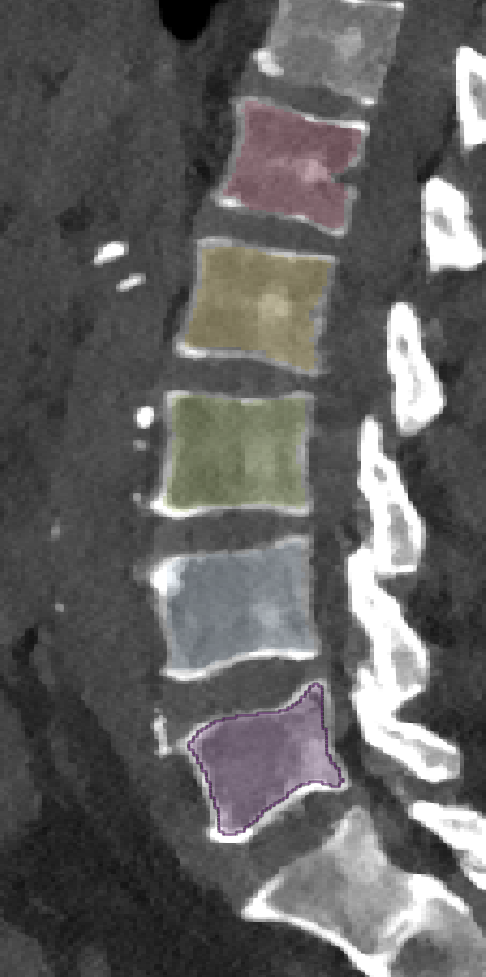}
            \caption{}
            \label{subfig:example1}
        \end{subfigure}
        \hfill
        \begin{subfigure}[b]{0.23\textwidth}
            \centering
            \includegraphics[width=\textwidth]{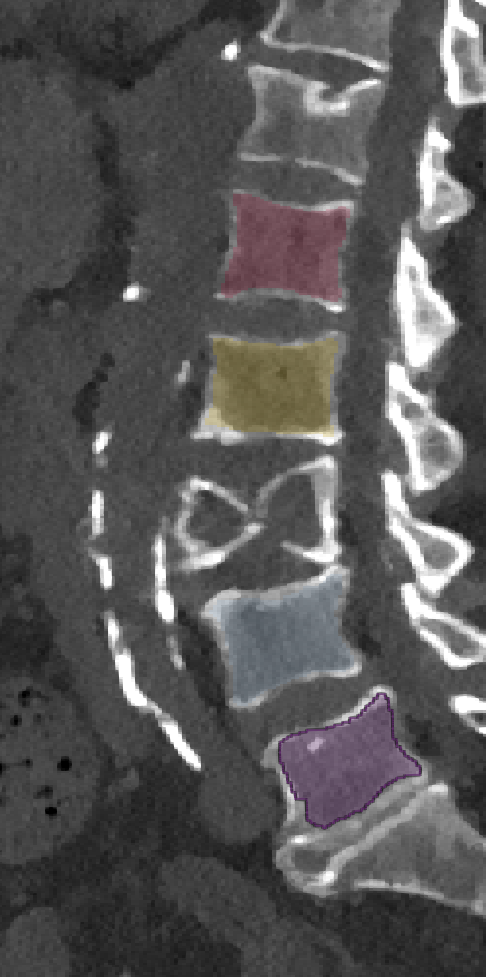}
            \caption{}
            \label{subfig:example2}
        \end{subfigure}
        \hfill
        \begin{subfigure}[b]{0.23\textwidth}
            \centering
            \includegraphics[width=\textwidth]{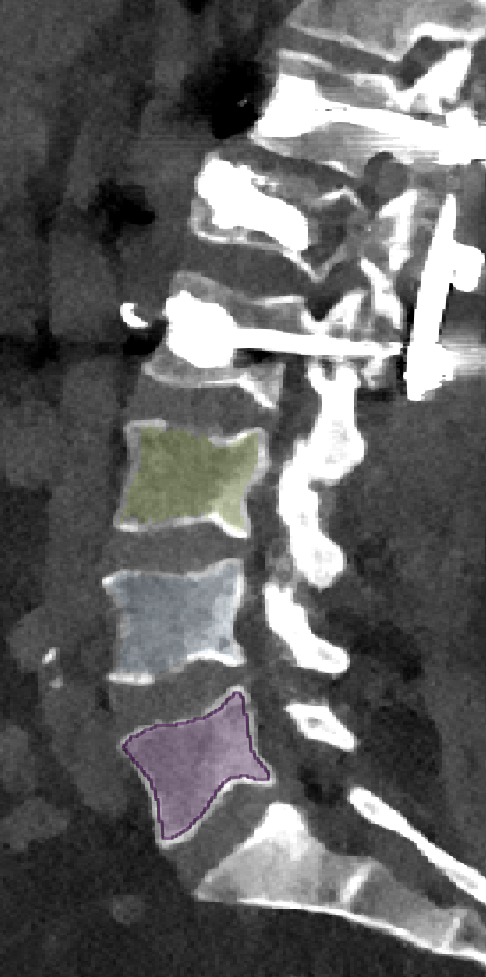}
            \caption{}
            \label{subfig:example3}
        \end{subfigure}
        \hfill
        \begin{subfigure}[b]{0.23\textwidth}
            \centering
            \includegraphics[width=\textwidth]{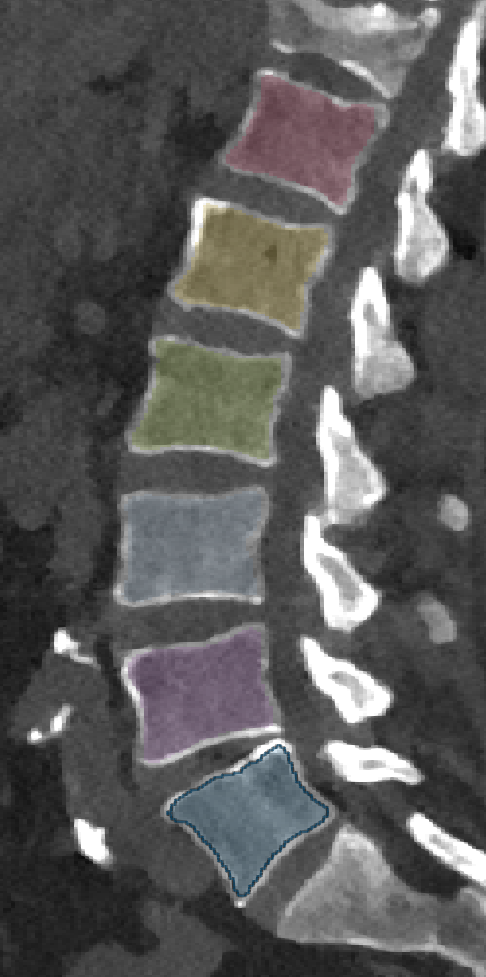}
            \caption{}
            \label{subfig:example4}
        \end{subfigure}

		\caption{Examples of data in sagittal view. Overlay of 40 keV Virtual Monoenergetic Image (VMI) with segmentation masks of lumbar trabecular bones. Each colour represents one lumbar vertebra. (\subref{subfig:example1}) complete lumbar spine with L1--L5 segmented, (\subref{subfig:example2}) compressed L3 wihtout segmentation mask, (\subref{subfig:example3}) metal implants in Th12--L2 without segmentation masks, (\subref{subfig:example4}) patient with lumbariastion -- segmentation available for L1--L6.}
        \label{fig:data_examples}
	\end{figure*}
    
    All examinations were performed using a dual-layer dual-energy CT scanner (Philips IQon Spectral CT). Whole-body low-dose CT scans were acquired from the skull base to the proximal tibia, with only the spinal region retained for the construction of the dataset.

    The acquisition parameters are summarized in Table~\ref{tab:acq_params}. All scans were acquired without intravenous contrast administration.
    
    \begin{table}[ht]
    \centering
    \caption{CT acquisition and reconstruction parameters.}
    \label{tab:acq_params}
    \begin{tabular}{ll}
    \hline
    Parameter & Value \\
    \hline
    Tube voltage & 120 kVp \\
    Tube current & 10 mA \\
    Matrix size & $512 \times 512$ \\
    Slice thickness & 0.9 mm \\
    Slice increment & 0.6 mm \\
    Reconstruction kernel & Sharp \\
    Reconstruction algorithm & iDose4 (level 4) \\
    Contrast agent & None \\
    In-plane pixel spacing (x–y) & 0.824–0.977 mm \\
    Slice spacing (z) & 0.6 mm \\
    \hline
    \end{tabular}
    \end{table}

    For each case study, multiple image series were reconstructed, including conventional CT, virtual monoenergetic images (40, 80, and 120 keV), and calcium-suppressed images (indices 25, 50, 75, and 100).

    All imaging data are provided in DICOM format. Segmentation masks are available in both NIfTI and DICOM-SEG formats. Clinical and acquisition metadata are provided in structured TSV files. The dataset follows a hierarchical directory structure at the patient-level to ensure reproducibility and traceability.

    All data were pseudonymized prior to release through removal of direct identifiers during DICOM export, and further anonymized as part of the TCIA curation process by defacing. No additional intensity normalization or preprocessing was applied beyond the original reconstruction pipelines.
    
    For this study, we used conventional CT series from the previously released \textit{Spinal-Multiple-Myeloma-SEG dataset (Version 1)} \citep{Nohel2026SpinalMM}. First, the images were cropped to the thoracolumbar region using the available vertebral segmentation masks. The region spanning vertebrae Th12 to L5 (or L6 when present) was extracted, with an additional margin of 20 voxels applied in all directions to preserve the anatomical context.

    All images were converted from DICOM to NIfTI format and reoriented to the RAS coordinate system. The initial segmentation masks of the trabecular compartments were predicted using a pretrained nnU-Net model \citep{curillova_2024_11245116, Isensee2021} publicly available on Zenodo. The model was originally trained on the LumVBCanSeg dataset \citep{Zhang2023} and is described in detail in the conference paper \citep{Nagyova2024}.

    The predicted segmentation masks were visually inspected and manually corrected by two biomedical engineers with experience in medical image analysis, if necessary. Manual corrections were performed using the ITK-SNAP software (version 4.0.2) \citep{Yushkevich2006}, with visual guidance from virtual monoenergetic images at 40 keV (VMI40), consistently used in all cases to improve the delineation of the trabecular bone. The trabecular compartment was defined within the vertebral body, with the posterior cortical border of the vertebral body serving as the posterior segmentation boundary. Posterior elements, including the pedicles and vertebral arch, were excluded from the segmentation masks. The correction process included verification of vertebral identification (L1--L5 / L6), removal of incorrect segmentation involving the Th12 vertebra, correction of regions where cortical bone had been included erroneously, and completion of missing trabecular regions. Cases containing extensive metal instrumentation in the lumbar spine that prevented reliable delineation of the trabecular compartment were excluded from the final dataset. Several examples of corrected segmentation masks are in Fig.~\ref{fig:data_examples}.

    All segmentation masks were reviewed by a board-certified radiologist with more than five years of clinical experience in musculoskeletal imaging. Additional refinements were performed if necessary, resulting in the final trabecular compartment segmentation masks provided with the dataset.

    All final segmentation masks were resampled and transformed back into the original CT image space. As a result, the released segmentation masks preserve the native image geometry, including image dimensions, voxel spacing, orientation, and spatial coordinates, allowing direct use within the corresponding DICOM image series.

    One case study was excluded from the final release because the lumbar spine region was not covered by the available CT acquisition. Consequently, trabecular compartment segmentation masks are available for 71 of the 72 studies included in the original Spinal-Multiple-Myeloma-SEG dataset.

    The primary limitation of the dataset is the presence of imaging artifacts caused by spinal instrumentation and severe degenerative changes, which can affect the quality of the segmentation. Vertebrae in which reliable delineation of the trabecular compartment was not feasible were excluded from the final release.

    This workflow was chosen to combine the efficiency of automated segmentation with the annotation quality required for public release and machine learning benchmarking.

\section{Validation}
    \subsection{Description of approaches ensuring quality of data}
    Quality assurance was performed throughout the entire data creation pipeline to ensure consistency and anatomical correctness of the released segmentation masks. The preprocessing steps, including the conversion from DICOM to NIfTI format, the reorientation to the RAS space and the cropping of the thoracolumbar region (Th12--L5 / L6), were visually inspected to verify correct spatial alignment and complete anatomical coverage. Initial segmentation masks of the trabecular compartment were generated using a pretrained nnU-Net model and subsequently evaluated for anatomical plausibility, with common error modes including incomplete trabecular coverage, inclusion of cortical bone, and occasional vertebral misidentification. All predicted segmentation masks were manually corrected by two biomedical engineers with experience in medical image analysis, focusing on vertebral labeling, removal of erroneous structures, and completion of missing regions. The final quality control was performed by a board-certified radiologist with more than five years of experience in musculoskeletal imaging, who verified anatomical consistency and performed additional refinements when necessary. In addition, consistency checks were conducted to ensure the correct correspondence between the imaging data, the segmentation masks and the patient identifiers, and one study was excluded due to incomplete coverage of the lumbar spine. As a result, segmentation masks of the trabecular compartments are available for 71 of the 72 studies included in the original dataset.
    
    \section{Conflicts of Interest}
    M.N., K.K., J.Ch., and R.J. collaborate with Philips Healthcare and have access to the Philips IntelliSpace Portal (version 12.1) under a research agreement. The remaining authors declare no competing interests.

\section{Acknowledgements}
    The paper and the research were supported by Philips Healthcare through the loan of an IntelliSpace Portal workstation.

    Computational resources were provided by the e-INFRA CZ project (ID:90254), supported by the Ministry of Education, Youth and Sports of the Czech Republic.

    Supported by Ministry of Health, Czech Republic - conceptual development of research organization (FNBr, 65269705)


\bibliography{sample}


	


\end{document}